\documentclass[runningheads]{llncs}
\usepackage{nicefrac}
\usepackage{amssymb}
\usepackage{booktabs}
\usepackage{graphicx}
\usepackage{color}
\usepackage{wrapfig}
\usepackage{here}
\usepackage{multirow}
\usepackage{tabularx}
\usepackage{url}
\usepackage{cite} 
\usepackage{comment}
\def\rnum#1{\resizebox{0.5em}{\height}{\expandafter{\romannumeral #1}}}

\begin{document}
\title{Automatic Segmentation, Localization, and Identification of Vertebrae in 3D CT Images Using Cascaded Convolutional Neural Networks}
\titlerunning{Automatic Segmentation, Localization and Identification of Vertebrae}
\author{Naoto Masuzawa\inst{1}   \and
        Yoshiro Kitamura\inst{1} \and
        Keigo Nakamura\inst{1}   \and
        Satoshi Iizuka\inst{2}   \and
        Edgar Simo-Serra\inst{3}
        }
\authorrunning{N. Masuzawa et al.}
\institute{Imaging Technology Center, Fujifilm Corporation, Minato, Tokyo, Japan\and
           Center for Artificial Intelligence Research, University of Tsukuba, Tsukuba, Ibaraki, Japan \and
           Department of Computer Science and Engineering, Waseda University, Shinjuku, Tokyo, Japan \\
           \email{naoto.masuzawa@fujifilm.com}
}
\maketitle
\begin{abstract}

This paper presents a method for automatic segmentation, localization, and identification of vertebrae in arbitrary 3D CT images.
Many previous works do not perform the three tasks simultaneously even though requiring a priori knowledge of which part of the anatomy is visible in the 3D CT images.
Our method tackles all these tasks in a single multi-stage framework without any assumptions.
In the first stage, we train a 3D Fully Convolutional Networks to find the bounding boxes of the cervical, thoracic, and lumbar vertebrae.
In the second stage, we train an iterative 3D Fully Convolutional Networks to segment individual vertebrae in the bounding box.
The input to the second networks have an auxiliary channel in addition to the 3D CT images.
Given the segmented vertebra regions in the auxiliary channel, the networks output the next vertebra. 
The proposed method is evaluated in terms of segmentation, localization, and identification accuracy with two public datasets of
15 3D CT images from the MICCAI CSI 2014 workshop challenge and 302 3D CT images with various pathologies introduced in \cite{glocker}.
Our method achieved a mean Dice score of 96\%, a mean localization error of 8.3 mm, and a mean identification rate of 84\%.
In summary, our method achieved better performance than all existing works in all the three metrics.

\keywords{Vertebrae \and Segmentation \and Localization \and Identification \and Convolutional neural networks}
\end{abstract}

\section{Introduction}
Automatic segmentation, localization, and identification of individual vertebrae from
3D CT (Computed Tomography) images play an important role in a pre-processing step of automatic analysis of the spine.
However, many previous works are not able to perform segmentation,
\begin{figure}[t]
    \centering
    \begin{tabular}{c}

        \begin{minipage}{0.45\hsize}
        \centering
        \includegraphics[bb=0 0 265 457, scale=0.37]{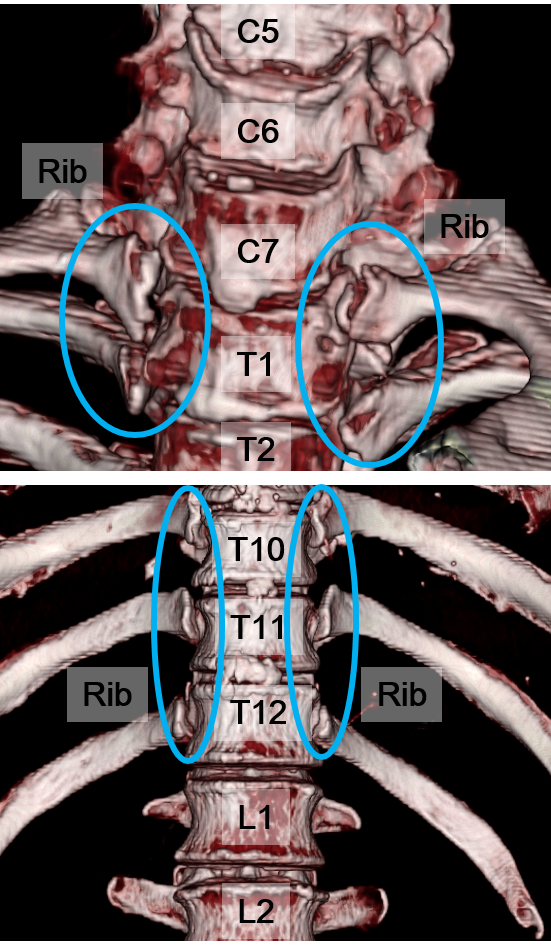}
        \vspace{-3mm}
        \caption{Differences in anatomy between cervical and thoracic vertebrae, and thoracic and lumbar vertebrae.}
        \label{fig:total_flow_2}
        \end{minipage}

        \begin{minipage}{0.05\hsize}
        \hspace{2mm}
        \end{minipage}

        \begin{minipage}{0.45\hsize}
            \centering
            \includegraphics[bb=0 0 447 575, scale=0.29]{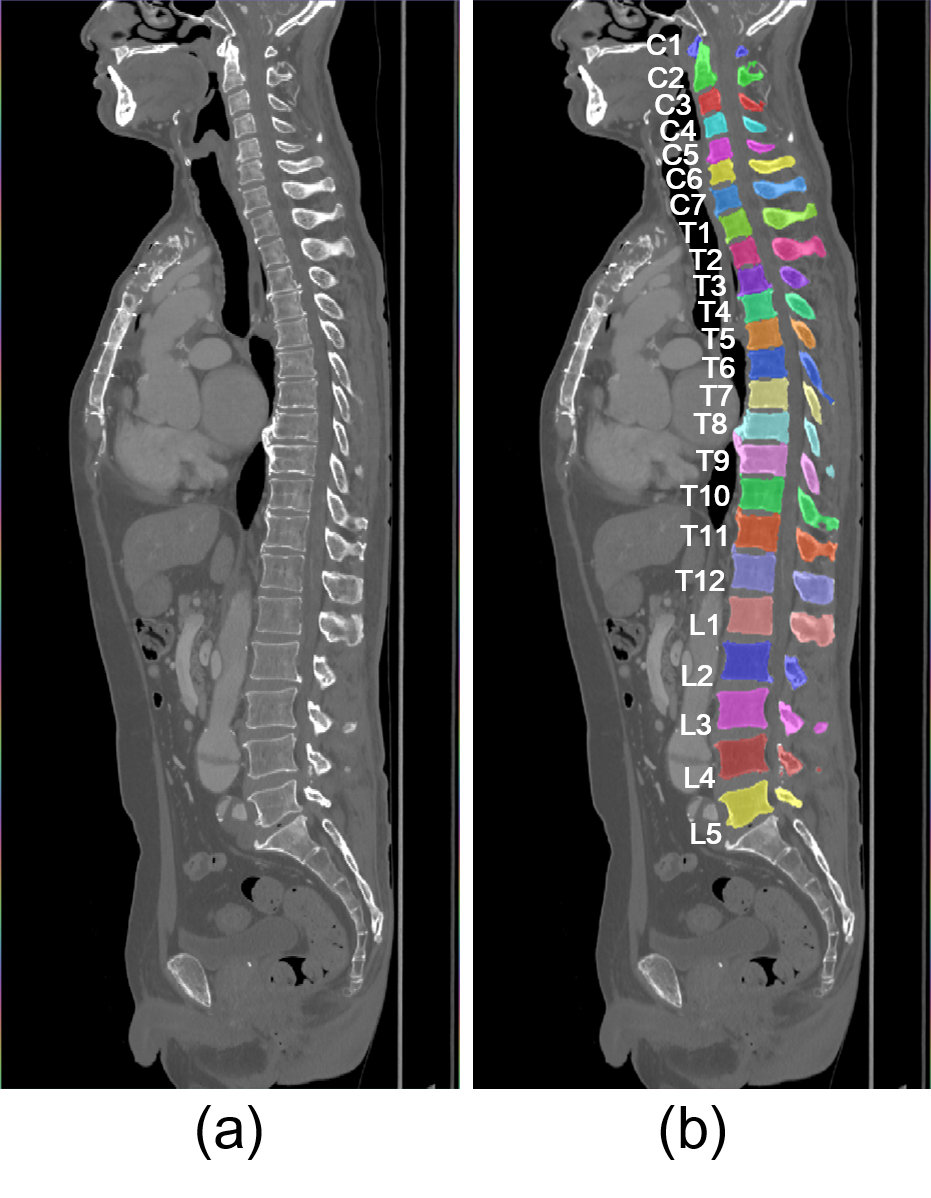}
            \vspace{-5mm}
            \caption{a) A sagittal slice of 3D CT images which includes cervical (C1-C7), thoracic (T1-T12), and lumbar (L1-L5) vertebrae.
                     b) Segmentation and identification of the individual vertebrae.}
            \label{fig:total_flow_2}
        \end{minipage}
    \end{tabular}
\end{figure}
localization, and identification simultaneously and require a priori knowledge of which part of the anatomy is visible in the 3D CT images.
\par We overcome those drawbacks with a single multi-stage framework.
More specifically, in the first stage, we train a 3D Fully Convolutional Networks (we call it "Semantic Segmentation Net"), which segments cervical, thoracic, and lumbar vertebrae 
so as to find the bounding boxes.
As shown in Figure 1, thoracic vertebrae are distinguished from the cervical and lumbar vertebrae by whether they connect to their ribs and therefore
it appears that the Semantic Segmentation Net performs well even if the field-of-view (FOV) is limited. In the second stage,
we train an iterative 3D Fully Convolutional Networks (we call it "Iterative Instance Segmentation Net"), which segments (i.e., predicts the labels of all voxels in the 3D CT images),
localizes (i.e., finds the centroids of all vertebrae), and identifies (i.e., assigns the anatomical labels) the vertebrae in the bounding box one-by-one.
Figure 2 shows an example input image and the corresponding image synthesized by the proposed method.
In summary, our contribution is as follows.
1) A two-stage coarse-to-fine approach for vertebrae segmentation, localization, and identification.
2) In-depth experiments and comparisons with existing approaches.

\section {Related work}
The challenges associated with automatic segmentation, localization, and identification of individual vertebrae are due to the following three points.
1) High similarity in appearance of the vertebrae.
2) The various pathologies such as the abnormal spine curvature and vertebral fractures.
3) The variability of input 3D CT images such as FOV, resolution, and image artifacts.
To address these challenges, many methods have been proposed.
Traditionally, vertebral segmentation has used mathematical methods such as atlas-based segmentation or deformable models \cite{related_1, related_2, related_3}.
Speaking of localization and identification, Glocker et al. \cite{glocker, glocker_2} proposed a method based on regression forests with a challenging dataset.
They introduced 302 3D CT images with various pathologies, the narrow FOV, and metal artifacts.
Recently, deep learning has been employed in the applications of vertebral segmentation, localization, and identification.
Yang et al. \cite{comp_2} proposed a deep image-to-image network (DI2IN) to predict centroid coordinates of vertebrae. 
On the other hand, the common way to segment vertebrae using deep learning is to use semantic segmentation
to predict the labels of all voxels in input 3D CT images. For example, Janssens et al. \cite{3d_fully} proposed a 3D fully convolutional neural networks (FCN) to segment lumbar vertebrae.
However, the way based on the semantic segmentation can segment vertebrae such as lumbar only when whole of the vertebrae is visible in 3D CT images.
This motivated Lessmann et al. \cite{nikolas_itr_net} to consider vertebral segmentation as an instance segmentation problem.
The networks introduced by Lessman et al. \cite{nikolas_itr_net} have an auxiliary channel in addition to the input.
Given the segmented vertebra regions in the auxiliary channel, the networks output the next vertebra.
Thus, the method proposed by Lessmann et al. \cite{nikolas_itr_net} is able to perform vertebral segmentation
even though whole of the vertebrae is not visible in 3D CT images and the number of vertebra is not known a priori.
\par Although the method by Lessmann et al. \cite{nikolas_itr_net} achieves high segmentation accuracy, it does not predict anatomical labels 
(i.e., cervical C1-C7, thoracic T1-T12, lumbar L1-L5) for each vertebra and it does not handle general 3D CT images where
it is not known in advance which part of the anatomy is visible. In fact, their method requires a priori knowledge of anatomy, such as lumbar 5.
On the other hand, our approach is able to predict anatomical labels and handle general 3D CT images.

\section{Proposed Method}
\begin{figure}[h]
    \centering
    \includegraphics[bb=0 0 1450 547, scale=0.2028]{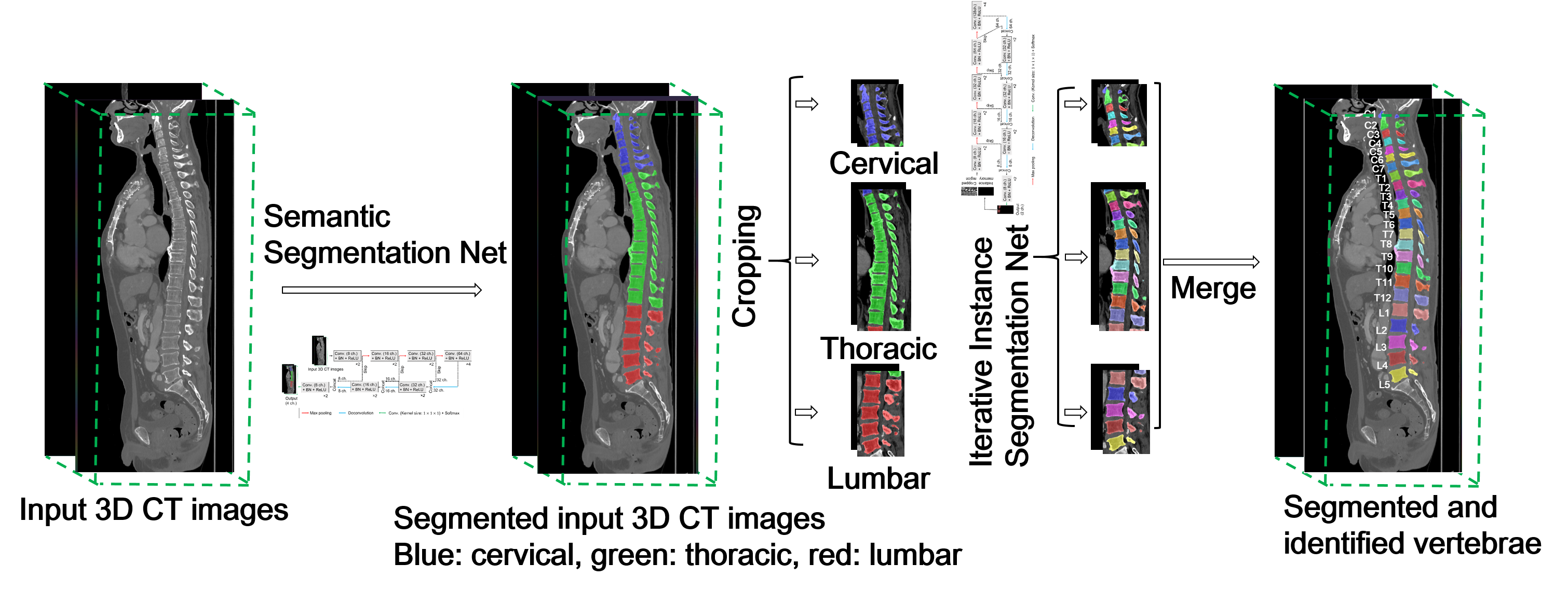}
    \vspace{-3mm}
    \caption{A schematic view of the present approach.}
    \label{fig:total_flow}
\end{figure}
Our method relies on a two-stage approach as shown in Figure 3.
The first stage aims to segment cervical, thoracic, and lumbar vertebrae from input 3D CT images.
Individual vertebrae are segmented in the second stage. Moreover, vertebral centroid coordinates and their labels are also obtained.
Below we first present our training dataset, followed by descriptions of the Semantic Segmentation Net and the Iterative Instance Segmentation Net.
\subsection{Training dataset}
We prepared 1035 3D CT images (head: 181, chest 477, abdomen: 270, leg: 107) for training
which are obtained from diverse manufacturer's equipment (e.g., GE, Siemens, Toshiba, etc.).
The leg 3D CT images were prepared for the purpose of suppressing false positive in the first stage.
The slice thickness ranges from 0.4 mm to 3.0 mm, and the in-plane resolution ranges from 0.34 mm to 0.97 mm.
They have been selected to contain the abnormal spine curvature, metal artifacts, and the narrow FOV.
Our spine model for training includes n = 25 individual vertebrae, where the regular 19 from the cervical, thoracic, and lumbar vertebrae consist irregular lumbar 6.
Reference segmentations of the visible vertebrae were generated by manually correcting automatic segmentations.

\subsection{Stage 1: Semantic Segmentation Net}
The convolutional neural networks are widely used to solve segmentation tasks in supervised learning technique.
Recent works have shown that this technique can be successfully applied to the multi-organ segmentation in 3D CT images \cite{multi_org}.
In our method, we develop the Semantic Segmentation Net which segment cervical, thoracic, and lumbar vertebrae from 3D CT images to find the bounding boxes.
\par Figure 4 shows a schematic drawing of the architecture.
Our architecture is based on a 3D FCN \cite{multi_org}.
For our Semantic Segmentation Net, the convolutions performed in each stage use volumetric kernels having size of 3$\times$3$\times$3 and strides of 1
followed by batch normalization \cite{batch_norm} and ReLU as the activate function,
the max pooling uses volumetric kernels having size of 2$\times$2$\times$2 and strides of 2, and the deconvolutions use volumetric kernels having size of 4$\times$4$\times$4 and strides of 2.
\begin{figure}[H]
    \centering
    \includegraphics[bb=0 0 1349 580, scale=0.238]{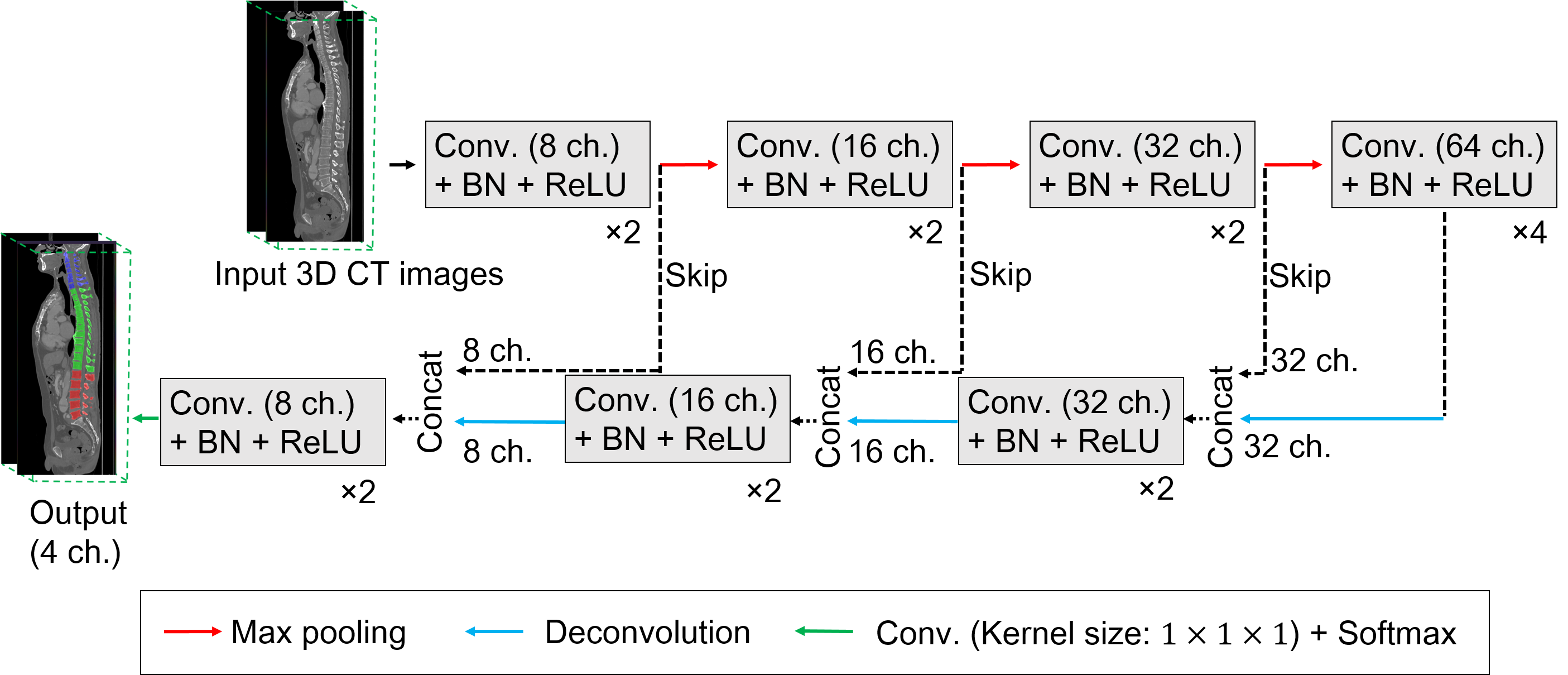}
    \vspace{-3mm}
    \caption{Architecture of the Semantic Segmentation Net.}
    \label{semantic}
\end{figure}

\subsubsection{Data augmentation and training}
In the preprocessing steps, input 3D CT images are clipped to the [-512.0, 1024.0] range and then normalized
to be in the [-1.0, 1.0] interval. After that, input 3D CT images are rescaled to 1.0 mm isotropic voxels.
For each training iteration, we randomly crop 160$\times$160$\times$160 voxels from the input 3D CT images
and apply data augmentation. In particular, we apply an affine transformation consisting of a random
rotation between -15 and +15 degrees, and random scaling between -20\% and +20\%, both sampled from uniform distributions.
In addition, we apply a Gaussian noise with $\mu$ = 0.0 and $\sigma$ = [0.0, 50.0/1536.0].
In the training iteration, bootstrapped cross entropy loss functions \cite{boot} were optimized with the Adam optimizer \cite{adam}
with a learning rate of 0.001 since the multi-class dice loss can be unstable.
The idea behind bootstrapping \cite{boot} is to backpropagate cross entropy loss not from all but a subset of voxels that 
the posterior probabilities are less than a threshold. In our experiment, 10\% of total voxels are used for the backpropagation.

\subsection{Stage 2: Iterative Instance Segmentation Net}

\begin{figure}[h]
    \centering
    \includegraphics[bb=0 0 1607 579, scale=0.216]{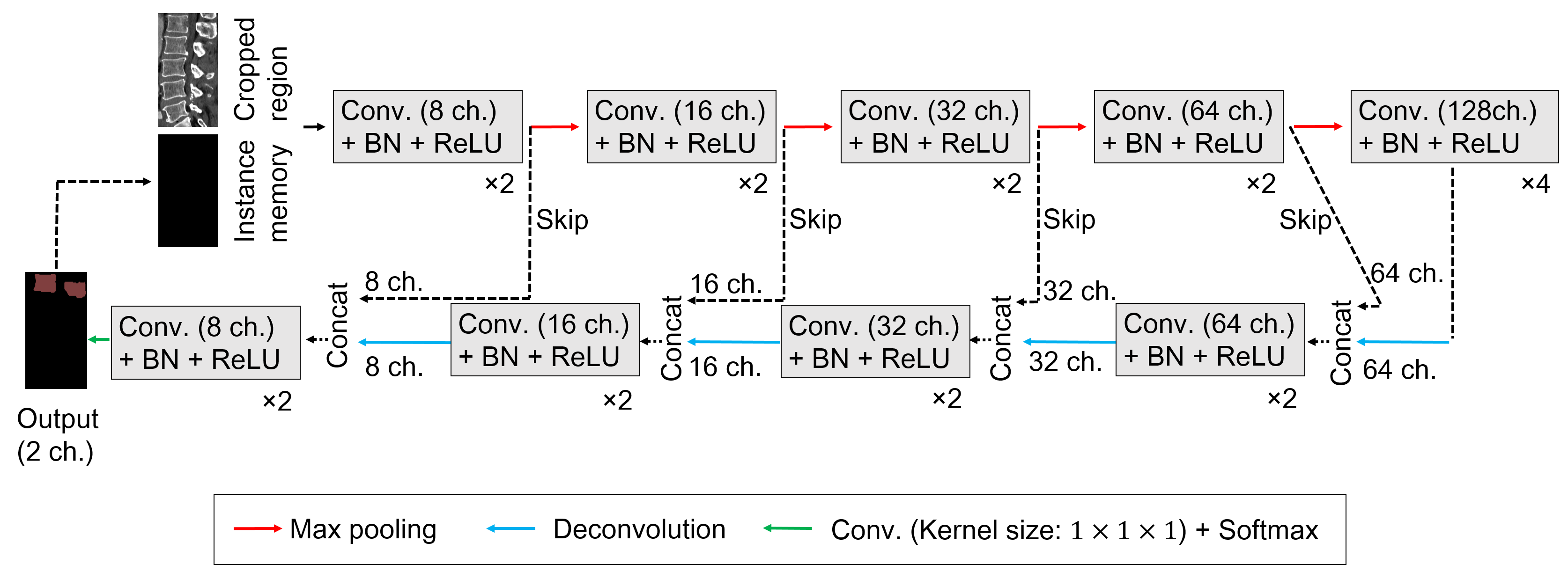}
    \vspace{0mm}
    \caption{Architecture of the Instance Segmentation Net.}
    \label{instance}
\end{figure}

The goal of the second stage is segmenting, localizing, and assigning anatomical labels to each vertebra.
To this end, we developed the Iterative Instance Segmentation Net inspired by Lessmann et al. \cite{nikolas_itr_net}.
The input to the Iterative Instance Segmentation Net has an auxiliary channel in addition to the 3D CT images.
Given the segmented vertebra regions in the auxiliary channel, the networks output the next vertebra.
The method by Lessmann et al. \cite{nikolas_itr_net} requires lumbar 5 region as a priori knowledge, and therefore it is not able to handle general 3D CT images.
By contrast, due to using the segmentation results in the first stage, our method is able to handle general 3D CT images.
\par Figure 5 shows a schematic drawing of the architecture.
For our Iterative Instance Segmentation Net, the convolutions performed in each stage use volumetric kernels having size of 3$\times$3$\times$3 and strides of 1
followed by batch normalization \cite{batch_norm} and ReLU as the activate function,
the max pooling uses volumetric kernels having size of 2$\times$2$\times$2 and strides of 2, and the deconvolutions use volumetric kernels having size of 4$\times$4$\times$4 and strides of 2.
The anatomical labels of individual vertebrae are counted starting from the boundaries of cervical and thoracic vertebrae or thoracic and lumbar vertebrae.
Finally, centroids of vertebrae are calculated using the segmentation results.

\subsubsection{Data augmentation and training}

In the preprocessing steps, similar to the first stage, input 3D CT images are clipped to the [-512.0, 1024.0] range and then normalized
to be in the [-1.0, 1.0] interval. After that, input 3D CT images are rescaled to 1.0 mm isotropic voxels.
For each training iteration, we randomly crop the spine region from the input 3D CT images and apply data augmentation.
In particular, we apply an affine transformation consisting of a random
rotation between -15 and +15 degrees, and random scaling between -20\% and +20\%, both sampled from uniform distributions.
In addition, we apply a Gaussian noise with $\mu$ = 0.0 and $\sigma$ = [0.0, 50.0/1536.0].
In the training iteration, the Dice loss of the segmented volume were optimized with the Adam optimizer \cite{adam} with a learning rate of 0.001.

\section{Experimental Results}

We present two sets of experimental results. The first one is on vertebral segmentation and the second one is about vertebral localization and identification.
We validate our algorithm with two public datasets of 15 3D CT images with reference segmentations from
the MICCAI CSI (Computational Spine Imaging) 2014 workshop challenge and 302 3D CT images of the patients with various types of pathologies introduced in \cite{glocker}.
There are unusual appearances in the second dataset such as abnormal spine curvature and metal artifacts.
In addition, the FOV of each volume varies widely.

\subsection{Segmentation Performance}

We evaluated our method in terms of the segmentation accuracy with the MICCAI CSI 2014 workshop challenge.
The CSI dataset consists of 15 3D CT images of healthy young adults, aged 20–34 years.
The images were scanned with either a Philips iCT 256 slice CT scanner or a Siemens Sensation 64 slice CT scanner (120 kVp, with IV-contrast).
The in-plane resolution ranges from 0.31 mm to 0.36 mm and the slice thickness ranges from 0.7 mm to 1.0 mm. Each volume cover thoracic and lumbar vertebrae.
We evaluate the segmentation performance using Average Symmetric Surface Distance (ASSD), Hausdorff Distance (HD), and Dice score
on condition that the final segmentation masks are rescaled to the resolution of the input 3D CT images.
The results on the CSI dataset is summarized in Table 1.
Our method achieved slightly better performance than existing methods. The examples of the segmentations and the anatomical labels obtained with our method are shown in Figure 6.
In all the 15 3D images, the Semantic Segmentation Net provided the Iterative Instance Segmentation Net with the accurate bounding boxes.
Moreover, the Iterative Instance Segmentation Net segmented the vertebrae precisely and predicted all of the anatomical labels.

\begin{table}[h]
    \caption{Comparison of Dice scores, ASSD and HD for segmentation results.}\label{tab1}
    \centering
    \setlength{\tabcolsep}{6pt}
    \vspace{-2mm}
    \begin{tabular}{rcccccc}
        \toprule
        Method & Dice score(\%) & ASSD (mm) & HD (mm)\\
        \midrule
        Janssens et al \cite{3d_fully} & 95.7 \% & 0.37 & 4.32\\
        Lessman et. al \cite{nikolas_itr_net} & 94.9 \% & 0.19 & -\\
        Our method & \bf96.6 \% & \bf0.10 & \bf2.11\\
        \bottomrule
    \end{tabular}
\end{table}

\begin{figure}[h]
    \centering
    \includegraphics[bb=0 0 691 491, scale=0.38]{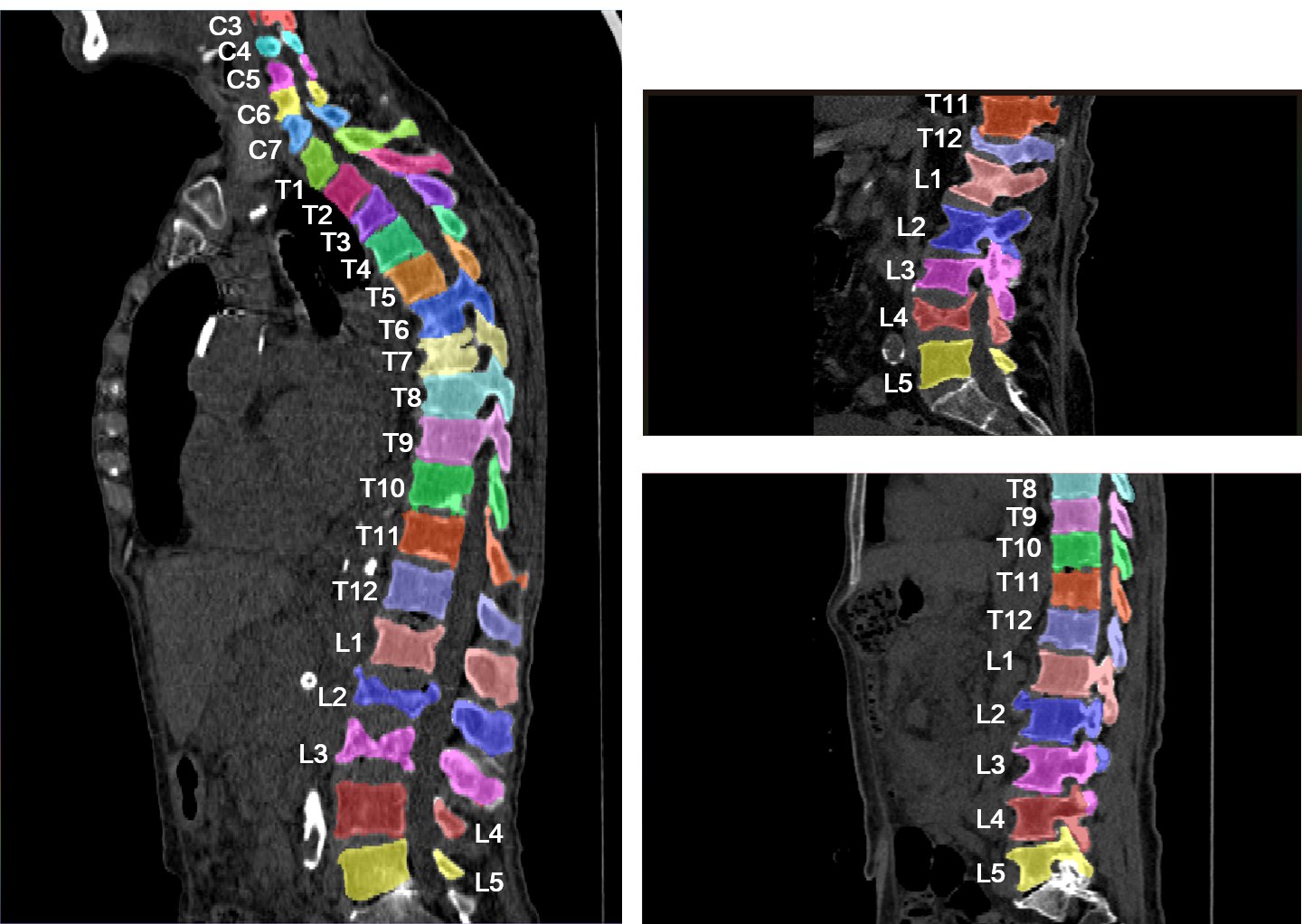}
    \vspace{-3mm}
    \caption{Segmentation results and predicted anatomical labels obtained with the proposed method.}
    \label{res_seg}
\end{figure}

\subsection{Identification and Localization Performance}

We evaluate localization and identification performance with 302 3D CT images introduced in \cite{glocker}.
This dataset is challenging since it includes wide varieties of anomalies such as the abnormal spine curvature and the metal artifacts.
Furthermore, the FOV of each volume is largely different.
In this dataset, the reference centroid coordinates of the vertebrae and the anatomical labels were given by clinical experts.
We evaluate our method with the two metrics described in \cite{glocker_2}, which are the Euclidean distance error (in mm) and identification rates (Id.Rates) defined in \cite{glocker}.
On calculating these metrics, the final segmentation masks are rescaled to the resolution of the input 3D CT images.
Table 2 shows a comparison between our method and previous works \cite{comp_1, comp_2}.
The mean localization error is 8.3 mm, and the mean identification rate is 84\%.
Our method achieved better performance than the other existing methods.

\begin{table}[]
    \caption{Comparison of localization errors in mm and identification rates.}\label{tab1}
    \centering
    \setlength{\tabcolsep}{4pt}
    \vspace{-1mm}
    \begin{tabular}{cccccccc}
        \toprule
        \multirow{2}{*}{}
        & Method & Mean & Std & Id.rates\\ \hline
        \multirow{6}{*}{All} & Glocker et al. \cite{glocker} & 12.4 & 11.2 & 70\% \\
        & Suzani et al. \cite{comp_1} & 18.2 & 11.4 & - \\
        & Yang et al.   \cite{comp_2} & 9.1 & \bf7.2 & 80\% \\
        & Yang et al.   \cite{comp_2} (+1000) & 8.5 & 7.7 & 83\% \\\
        & Our method & \bf8.3 & 7.6 & \bf84\% \\ \hline
        \multirow{6}{*}{Cervical} & Glocker et al. \cite{glocker} & 7.0 & 4.7 & 80\% \\
        & Suzani et al. \cite{comp_1} & 17.1 & 8.7 & - \\
        & Yang et al.   \cite{comp_2} & 6.6 & 3.9 & 83\% \\
        & Yang et al.   \cite{comp_2} (+1000) & 5.8 & 3.9 & 88\% \\
        & Our method & \bf5.7 & \bf3.8 & \bf89\% \\ \hline
        \multirow{6}{*}{Thoracic} & Glocker et al. \cite{glocker} & 13.8 & 11.8 & 62\%\\
        & Suzani et al. \cite{comp_1} & 17.2 & 11.8 & - \\
        & Yang et al.   \cite{comp_2} & 9.9 & 7.5 & 74\% \\
        & Yang et al.   \cite{comp_2} (+1000) & 9.5 & 8.5 & 78\% \\
        & Our method & \bf9.3 & \bf8.3 & \bf79\% \\ \hline
        \multirow{6}{*}{Lumbar} & Glocker et al. \cite{glocker} & 14.3 & 12.3 & 75\% \\
        & Suzani et al. \cite{comp_1} & 20.3 & 12.2 & - \\
        & Yang et al.   \cite{comp_2} & 10.9 & 9.1 & 80\% \\
        & Yang et al.   \cite{comp_2} (+1000) & 9.9 & 9.1 & 84\% \\
        & Our method & \bf9.8 & \bf9.0 & \bf85\% \\
        \bottomrule
    \end{tabular}
\end{table}

\section{Conclusion}

In this paper, we propose a multi-stage framework for segmentation, localization and identification of vertebrae in 3D CT images.
A novelty of this framework is to divide the three tasks into two stages.
The first stage is multi-class segmentation of cervical, thoracic, and lumbar vertebrae.
The second stage is iterative instance segmentation of individual vertebrae.
By doing this, the method successfully works without a priori knowledge of which part of the anatomy is visible in the 3D CT images.
This means that the method can be applied to a wide range of 3D CT images and applications.
In the experiments using two public datasets, the method achieved the best Dice score for volume segmentation,
and achieved the best mean localization error and identification rate.
As far as we know, this is the first unified framework that tackles the three tasks simultaneously with the state of the art performance.
We hope that the proposed method will help doctors in clinical practice.

\bibliography{ref.bib}

\end{document}